# A Categorical View on Algebraic Lattices in Formal Concept Analysis

Pascal Hitzler[1]
Markus Krötzsch[2]
Guo-Qiang Zhang[3]

[1] Institut AIFB, Universität Karlsruhe, Germany.
[2] Fakultät für Informatik, Technische Universität Dresden, Germany.
[3] Department of Electrical Engineering and Computer Science, Case Western Reserve University, Cleveland, Ohio, U.S.A.

**Abstract**

Formal concept analysis has grown from a new branch of the mathematical field of lattice theory to a widely recognized tool in Computer Science and elsewhere. In order to fully benefit from this theory, we believe that it can be enriched with notions such as *approximation by computation* or *representability*. The latter are commonly studied in denotational semantics and domain theory and captured most prominently by the notion of *algebraicity*, e.g. of lattices. In this paper, we explore the notion of algebraicity in formal concept analysis from a category-theoretical perspective. To this end, we build on the the notion of *approximable concept* with a suitable category and show that the latter is equivalent to the category of algebraic lattices. At the same time, the paper provides a relatively comprehensive account of the representation theory of algebraic lattices in the framework of Stone duality, relating well-known structures such as Scott information systems with further formalisms from logic, topology, domains and lattice theory.

# 1 Introduction

Algebraic lattices conveniently represent computationally relevant properties. As partial orders they allow for the expression of amounts of *information content*. Distinguished elements — called *compact or finite* — stand for computationally representable information. Every element or information item not directly representable can be approximated by representable, i.e. compact, items.

So algebraic lattices can be identified as computationally relevant structures, and as such have found applications in Computer Science, most prominently in the theory of denotational semantics, domain theory (see, e.g. [AJ94]), but recently also in aspects regarding knowledge representation and reasoning (see e.g. [RZ01, ZR04, Hit04]).

As can be expected from rich mathematical structures such as algebraic lattices, a multitude of possible characterizations have been established, ranging from the classical correspondence between algebraic lattices and their semilattices of compacts [GHK$^+$03], over logical characterizations such as Scott information systems [Sco82a], to topological investigations via the Scott topology [Joh82, Abr91]. Following Abramsky's programme of *domain theory in logical form*, each of these representations is associated with either the *spacial* or the *localic* side of Stone duality: the former includes syntactical, logical, and axiomatic formalisms, while the latter typically incorporates semantical, observational, and denotational aspects. The equivalence of both worlds leads to rather pleasant results of soundness and completeness of corresponding proof systems and model theories.

We add to this collection by a representation of algebraic lattices based on the framework of formal concept analysis (FCA, [GW99]). Originally, FCA was conceived as an alternative formulation of the theory of complete lattices, motivated by philosophical considerations [Wil82]. In the meantime, FCA has grown from a new branch of lattice theory to a widely recognized tool in Computer Science (see, e.g., [Stu02]). Prominent applications concern areas such as Data- and Textmining, Knowledge Representation and Reasoning, Semantic Web, Computational Linguistics. FCA starts from *formal contexts*, syntactical descriptions of object-attribute relations, and lifts them to closure operators and complete lattices. While this suggests a logical viewpoint based on the given (deductive) closure, the derived logical entailment lacks the important property of compactness: some conclusions can only be drawn from infinite sets of premises [Zha03a]. This motivates a deviation from the classical definition of closures in FCA to ensure Scott continuity of the derived closure operators (the so called algebraic ones), thus recovering compactness and switching to complete lattices that are algebraic. We achieve this by introducing the (complete algebraic) lattice of *approximable concepts* [ZS0x], obtained from given object-attribute relations analogous to the classical construction used in FCA, but at the same time conforming to the insights concerning computationally relevant structures as studied in domain theory.

The strong interest in algebraic lattices indeed stems only in part from the appealing way in which these structures capture the possibility of approximating infinite computation by finite elements. The full strength of the theory only becomes apparent when Scott continuous functions are employed as morphisms of a category **Alg** of algebraic lattices. The interplay between the lattices and these morphisms is highly satisfactory: the set of all Scott continuous functions between two algebraic lattices can again be viewed as an algebraic lattice, and **Alg** is in fact cartesian closed. Consequently, we augment the above characterizations of algebraic lattices by suitable notions of morphisms, inducing in each case a category that is equivalent to **Alg**. We thus obtain a cartesian closed category of formal contexts corresponding to the new notion of approximable concepts.

At the same time, this article gives a relatively comprehensive account of the numerous representations of algebraic lattices by offering a fresh, unified, and largely self-contained treatment of the theory. The new approach via FCA provides additional insights into the nature of the well-known formalisms. In particular, we give a direct proof of the cartesian closedness of the new category of formal contexts, hence obtaining novel categorical product and function space constructions based on formal contexts. Particularly, the formulation of function spaceenhances our understanding of *approximable mappings*, the class of morphisms Scott conceived for his information systems [Sco82a]. Indeed, these relations turn out to be immediate descriptions of sets of step functions, sufficient to capture all Scott continuous functions between the corresponding algebraic lattices.

Our discussion will also expose the connections between algebraic lattices and the conjunctive fragment of propositional logic — an approach that appears to be rather intuitive from the viewpoint of Computer Science and also brings to bear on the results from [HW03, Hit04]. As encompassed in [Abr91], this is achieved through the Lindenbaum algebras of these logics. Our profit, besides finding a simple access to Scott information systems, is an alternative view on approximable mappings as *multilingual sequent calculi*, as considered in [JKM99] for more expressive logics.

The structure of this paper is as follows. In the next Section 2 the most fundamental definitions from order theory, topology, and category theory are recalled. Section 3 starts the discussion of algebraic lattices from a domain theoretic perspective, with special emphasis on the role of the semilattice of compact elements. Thereafter, Section 4 introduces appropriate notions of morphisms for such semilattices, which are shown to be equivalent to Scott continuous functions between the corresponding algebraic lattices. Section 5 then introduces a category of formal contexts equivalent to the algebraic lattices and Scott continuous functions, and gives an explicit proof of the cartesian closure of this new category. Building on the prototypical categorical equivalences established earlier, Section 6 introduces further representation theorems from logic and topology, which are then connected using Stone duality. Finally, Section 7 gives pointers to further literature and hints at possible extensions of given results.

A very preliminary report on some of the results in this paper has appeared as [HZ04]. The notion of *approximable concept* has first been proposed in [ZS0x], but without exploring its category-theoretical content.

*Acknowledgements.* The first author acknowledges support by the German Ministry of Education and Research under the SmartWeb project, and by the European Union under the KnowledgeWeb network of excellence. The second author gratefully acknowledges support by *Case Western Reserve University*, Cleveland/Ohio, where most of his work was carried out, and sponsorship by the *German Academic Exchange Service* (DAAD) and by the *Gesellschaft von Freunden und Förderern der TU Dresden e.V.*

## 2 Preliminaries and Notation

We first give some basic definitions of order theory at least *to fix notations*. Our main reference will be [GHK[+]03]. A more gentle first introduction is given in [DP02].

A partially ordered set (poset) is a set $P$ with a reflexive, symmetric, transitive relation $\leq \subseteq P \times P$. If $(P, \leq)$ is a poset, then its dual is the poset $(P, \geq)$. We denote posets by their carrier set as long as the partial order is clear from the context.

**Definition 2.1** Consider a poset $L$. A non-empty subset $D \subseteq L$ is *directed* if, for any $x, y \in D$, there is some element $z \in D$ such that $x \leq z$ and $y \leq z$. If every directed subset $D \subseteq L$ has a least upper bound (supremum, join) $\bigvee D$ in $L$, then $L$ is a *directed complete partial order* (dcpo).

$L$ is a *complete lattice* if every subset $S \subseteq L$ has a least upper bound $\bigvee S$ and a greatest lower bound (infimum, meet) $\bigwedge S$. For a set $X$, $2^X$ denotes the powerset lattice, i.e. the complete lattice of all subsets of $X$ under inclusion.

We recall that a poset that has all infima also has all suprema, and vice versa, so that one of these conditions is in fact sufficient. Furthermore we fix some basic terminologies about lattices.

**Definition 2.2** A poset $L$ is a *lattice* if every two elements of $L$ have a supremum and an infimum. These meets and joins of binary sets will be written in infix: $\bigvee\{x, y\} = x \vee y$ and $\bigwedge\{x, y\} = x \wedge y$. $L$ is distributive if, for all $x, y, z \in L$, one finds $x \wedge (y \vee z) = (x \vee y) \wedge (x \vee z)$.

An element $x \in L$ is called

- *meet-irreducible* if $y \wedge z = x$ implies $y = x$ or $z = x$,

- *meet-prime* if $y \wedge z \leq x$ implies $y \leq x$ or $z \leq x$.

Join-irreducible and join-prime elements are defined dually.

In a distributive lattice, the meet-irreducibles are exactly the meet-primes, and this will be the only case considered in this paper. Furthermore we want to talk about functions between partially ordered sets.

**Definition 2.3** Consider posets $P$ and $Q$, and a function $f : P \to Q$. Then $f$ is *monotone* if it preserves the order of $P$, i.e. $x \leq y$ in $P$ implies $f(x) \leq f(y)$ in $Q$. Moreover, *f preserves (directed) suprema* if, for any (directed) $S \subseteq P$ such that $\bigvee S$ exists, one finds that $\bigvee f(S) = \bigvee \{f(s) \mid s \in S\}$ exists and that $f(\bigvee S) = \bigvee f(S)$. Preservation of infima is defined dually. A function that preserves directed suprema is also called *Scott continuous*. An *order-isomorphism* is a bijective monotone function that has a monotone inverse.

Note that preservation of directed suprema (infima) always entails monotonicity, since every pair of elements $x \leq y$ induces a directed set $\{x, y\}$ for which preservation of suprema (infima) implies $f(x) \leq f(y)$ as required.

We also need a little general topology. Our view on topology largely agrees with [Smy92].

**Definition 2.4** Let $X$ be a set and let $\tau \subseteq 2^X$ be a system of subsets of $X$. $(X, \tau)$ is a *topological space* if $\tau$ contains $X$ and the empty set, and is closed under arbitrary unions and finite intersections. The members of such a system $\tau$ are called *open sets* and the complete lattice $(\tau, \subseteq)$ is called the *open set lattice*. The complements of open sets are the *closed sets*. If confusion is unlikely, we will denote topological spaces by their sets of points. For a topological space $X$, we also use $\Omega(X)$ to denote its open set lattice.

A subset $B$ of $\tau$ is a *base* of $\tau$ if every open set is equal to the union of all members of $B$ it contains.

The appropriate mappings between topological spaces are *continuous functions*.

**Definition 2.5** Consider topological spaces $X$ and $Y$, and a function $f : X \to Y$. Then $f$ is *continuous* if its inverse image preserves open sets, i.e. for every open set $O \subseteq Y$, the set $f^{-1}(O) = \{x \in X \mid f(x) \in O\}$ is open in $X$. If $f$ is bijective and both $f$ and $f^{-1}$ are continuous then $f$ is a *homeomorphism*. The topological spaces $X$ and $Y$ are said to be *homeomorphic* if a homeomorphism between them exists.

Finally, a minimum amount of category theory is utilized in this paper, in order to present relationships of the different concepts to their full extent. Our terminology follows [Bor94]. Other good references include [Mac71], and the more easy-paced introductions [LR03] and [McL92]. A *category* **C** consists of

(i) a class $|\mathbf{C}|$ of *objects* of the category,

(ii) for all $A, B \in |\mathbf{C}|$, a set $\mathbf{C}(A, B)$ of *morphisms* from $A$ to $B$,

(iii) for all $A, B, C \in |\mathbf{C}|$, a composition operation
$$\circ : \mathbf{C}(B, C) \times \mathbf{C}(A, B) \to \mathbf{C}(A, C),$$

(iv) for all $A \in |\mathbf{C}|$, an *identity morphism* $\mathrm{id}_A \in \mathbf{C}(A, A)$,

such that for all $f \in \mathbf{C}(A, B)$, $g \in \mathbf{C}(B, C)$, $h \in \mathbf{C}(C, D)$, the associativity axiom $h \circ (g \circ f) = (h \circ g) \circ f$, and the identity axioms $\mathrm{id}_B \circ f = f$ and $g \circ \mathrm{id}_B = g$ are satisfied. As usual, we write $f : A \to B$ for morphisms $f \in \mathbf{C}(A, B)$. The *opposite* $\mathbf{C}^{\mathrm{op}}$ of a category $\mathbf{C}$ is defined by setting $|\mathbf{C}^{\mathrm{op}}| = |\mathbf{C}|$ and $\mathbf{C}^{\mathrm{op}}(A, B) = \mathbf{C}(B, A)$. A morphism $f : A \to A'$ is an *isomorphism*, if it has an inverse, i.e. if there is a (necessarily unique) morphism $g : A' \to A$ with $g \circ f = \mathrm{id}_A$ and $f \circ g = \mathrm{id}_{A'}$.

A *functor* $\mathsf{F}$ from a category $\mathbf{A}$ to a category $\mathbf{B}$ consists of

(i) a mapping $|\mathbf{A}| \to |\mathbf{B}|$ of objects, where the image of an object $A \in |\mathbf{A}|$ is denoted by $\mathsf{F}A$,

(ii) for all $A, A' \in |\mathbf{A}|$, a mapping $\mathbf{A}(A, A') \to \mathbf{B}(\mathsf{F}A, \mathsf{F}A')$, where the image of a morphism $f \in \mathbf{A}(A, A')$ is denoted by $\mathsf{F}f$, such that for all $A, B, C \in |\mathbf{A}|$ and all $f \in \mathbf{A}(A, B)$ and $g \in \mathbf{A}(B, C)$ we have $\mathsf{F}(f \circ g) = \mathsf{F}f \circ \mathsf{F}g$ and $\mathsf{F}\,\mathrm{id}_A = \mathrm{id}_{\mathsf{F}A}$.

The third basic ingredient of category theory are *natural transformations*. Given two functors $\mathsf{F}, \mathsf{G} : \mathbf{A} \to \mathbf{B}$, a family of morphisms $\eta = (\eta_A : \mathsf{F}A \to \mathsf{G}A)_{A \in |\mathbf{A}|}$ is a natural transformation from $\mathsf{F}$ to $\mathsf{G}$, if, for all morphisms $f : A \to A'$ of $\mathbf{A}$, one has that $\eta_{A'} \circ \mathsf{F}f = \mathsf{G}f \circ \eta_A$. This situation is denoted by $\eta : \mathbf{A} \Rightarrow \mathbf{B}$. A natural transformation $(\eta_A)_{A \in |\mathbf{A}|}$ is a *natural isomorphism* if all of its members are isomorphisms.

More specific notions will be introduced as they are needed.

## 3 Algebraic lattices

In this section we introduce algebraic lattices and review their most well-known characterizations in terms of the sub-poset of compact elements and closure systems of Scott continuous closure operators. The material basically follows [GHK+03], to which we refer for the details of the proofs which we omit to avoid replication. We start with a basic definition.

**Definition 3.1** Consider a dcpo $P$. An element $c \in P$ is *compact* if, for every directed set $D \subseteq P$ we have that $c \leq \bigvee D$ implies $c \leq d$ for some $d \in D$. The set of all compact elements of $P$ is denoted by $\mathsf{K}(P)$. We usually consider $\mathsf{K}(P)$ to be a sub-poset of $P$.

We note the following

**Proposition 3.2** Let $L$ be a complete lattice with compact elements $a, b \in \mathsf{K}(L)$ and least element $\bot$. Then $a \vee b$ and $\bot$ are compact.

Proposition 3.2 contains important information about the structure of the sub-poset of compact elements of a complete lattice. The following definition makes the properties of $\mathsf{K}(L)$ precise.

**Definition 3.3** A poset $S$ is a *join-semilattice*, if any two elements $a$, $b$ in $S$ have a least upper bound $a \vee b$. Dually, in a meet-semilattice any two elements have a greatest lower bound.

We conclude that the poset $\mathsf{K}(L)$ of compact elements of a complete lattice is a join-semilattice with least element under the order of $L$. However, for a full characterization we shall also be interested in the opposite direction, i.e. given a join-semilattice, we would like to construct a complete lattice. The right tool for this endeavor is that of *ideal completion*, introduced next. Given a set $X$ we define $\downarrow X = \{y \mid \text{there is } x \in X \text{ such that } y \leq x\}$ and $\uparrow X = \{y \mid \text{there is } x \in X \text{ such that } x \leq y\}$; a set is called an *upper* (respectively, *lower*) set if $X = \uparrow X$ (respectively, $X = \downarrow X$). Upper and lower sets of singleton sets $\{x\}$ are denoted by $\uparrow x$ and $\downarrow x$, respectively.

**Definition 3.4** Consider a partially ordered set $P$. A subset $I \subseteq P$ is an *ideal* if it is a directed lower set. The *ideal completion* $\mathsf{Idl}(P)$ is the collection of all ideals of $P$ partially ordered via subset inclusion.

Note that lower sets $\downarrow x$ are always ideals — the *principle ideals* generated by the element $x$. On the other hand, the empty set is not an ideal, since directed sets need to be non-empty. We see below that the ideal completion of any join-semilattice with least element is a complete lattice. However, not all complete lattices arise in this way. The next definition provides the appropriate characterization.

**Definition 3.5** A complete lattice $L$ is an *algebraic lattice*, if for every element $x \in L$, we have $x = \bigvee (\downarrow x \cap \mathsf{K}(L))$.

One can easily see from Proposition 3.2 that any set of the form $\downarrow x \cap \mathsf{K}(L)$ is necessarily directed. Now we are ready to state the important

**Theorem 3.6 ([GHK$^+$03] Proposition I-4.10)** Let $L$ be an algebraic lattice and let $S$ be a join-semilattice with least element.

(i) $\mathsf{K}(L)$ is a join-semilattice with least element, where the order is induced by that given on $L$.

(ii) $\mathsf{Idl}(S)$ is a an algebraic lattice, where join is given by set-intersection.

(iii) $S$ is order-isomorphic to $\mathsf{K}(\mathsf{Idl}(S))$ via the isomorphism $f : S \to \mathsf{K}(\mathsf{Idl}(S))$ : $a \mapsto \downarrow a$.

(iv) $L$ is order-isomorphic to $\mathsf{Idl}(\mathsf{K}(L))$ via the isomorphism $g : L \to \mathsf{Idl}(\mathsf{K}(L))$ : $x \mapsto \downarrow x \cap \mathsf{K}(L)$.

This result demonstrates that we can represent any algebraic lattice — up to isomorphism — by an appropriate semilattice and vice versa. We subsequently obtain a number of alternative characterizations from this statement and its proof. A first observation is that Theorem 3.6 assures that every algebraic lattice is isomorphic to a lattice of sets. More precisely, for an algebraic lattice $L$, we established an isomorphism to a subset of the powerset of its compact elements $2^{\mathsf{K}(L)}$. Now one may ask how to characterize those substructures of powersets which yield algebraic lattices. The tool for this purpose are closure operators.

**Definition 3.7** Consider a poset $P$ and a function $c : P \to P$. Then $c$ is a *closure operator* if the following hold for all elements $x, y \in P$

(i) $c(x) = c(c(x))$ ($c$ is idempotent)

(ii) $x \leq c(x)$ ($c$ is inflationary)

(iii) $x \leq y$ implies $c(x) \leq c(y)$ ($c$ is monotone)

An important result about this kind of operators is that they can be characterized completely by their images, the *closure systems*. Explicitly, we have the following.

**Proposition 3.8 ([GHK$^+$03] Proposition O-3.13)** Let $L$ be a complete lattice and let $c$ be a closure operator on $L$. Then $c$ preserves arbitrary infima. Especially, its image $c(L) = \{c(x) \mid x \in L\}$ is closed under arbitrary infima in $L$. Conversely, any subset $C$ of $L$ that is closed under arbitrary infima in $L$ induces a unique closure operator $c$ with image $C$, given by $c : L \to L : x \mapsto \bigwedge\{y \in C \mid x \leq y\}$.

In Theorem 3.6(ii) it was shown that the set of ideals is closed under arbitrary intersections. By the above proposition this assures that $\mathsf{Idl}(S)$ is a closure system on $2^S$, which can be uniquely characterized by a closure operator. However, not every closure system is algebraic, such that a further restriction on the class of closure operators is required. It turns out that Scott continuity (see Definition 2.3) is what is needed to further extend the representation of algebraic lattices.

**Theorem 3.9 ([GHK$^+$03] Corollary I-4.14)** Any algebraic lattice $L$ is isomorphic to the image of a Scott continuous closure operator on the powerset $2^{K(L)}$. The operator is given by assigning to any set of compacts the least ideal which contains this set. Conversely, the image of any such closure is an algebraic lattice, where the compacts are exactly the images of finite sets of compacts.

This gives us a third characterization of algebraic lattices. One is tempted to develop a similar statement for join-semilattices with least element. Indeed, any closure operator on the semilattice of finite elements of a powerset can uniquely be extended to a Scott continuous closure on the powerset. However, it is not true that all join-semilattices are images of closure operators on the semilattice of finite subsets of some set. This is easy to see by noting that any collection of finite sets can only have finite descending chains, i.e. it satisfies the *descending chain condition* (see [DP02]). Yet there are join-semilattices with least element that do not have this property, like for example the non-negative rational numbers in their natural order. What we can say is the following.

**Corollary 3.10** For any join-semilattice $S$ with least element, there is a closure operator $c : 2^S \to 2^S$, such that $S$ is isomorphic to the image of the finite elements of $2^S$ under $c$. Conversely, the finite-set image of any closure operator on a powerset is a join-semilattice with least element.

**Proof.** Note that any closure operator $c$ on a powerset induces a unique Scott continuous closure $c'$ by setting $c'(X) = \bigcup\{c(A) \mid A \subseteq X,\ A$ finite$\}$, where $c'$ agrees with $c$ on all finite sets. Then combine Theorems 3.6 and 3.9, especially the characterization of compact closed subsets. □

The significance of this statement will become apparent in Section 5.

# 4 Approximable mappings

So far we have only provided object-level correspondences between algebraic lattices and join-semilattices. We supplement this with suitable morphisms which turn these relations into an equivalence of the respective categories. On the side of algebraic lattices, one typically employs Scott continuous functions to form a category **Alg**. This definition leads to a rather advantageous property, namely *cartesian closedness*, which will be discussed in the next section. The aim of this section is to identify a notion of morphism for join-semilattices that produces a category which is *equivalent* to **Alg**.

**Definition 4.1** Consider categories **A** and **B**. An *equivalence of categories* **A** and **B** is constituted by a pair of functors $\mathsf{F} : \mathbf{A} \to \mathbf{B}$ and $\mathsf{G} : \mathbf{B} \to \mathbf{A}$, together with a pair of natural isomorphisms $\eta : \mathsf{GF} \Rightarrow \mathrm{id}_\mathbf{A}$ and $\epsilon : \mathsf{FG} \Rightarrow \mathrm{id}_\mathbf{B}$, where $\mathrm{id}_\mathbf{A}$ and $\mathrm{id}_\mathbf{B}$ denote the identity functors on the respective categories.

It is well-known that a functor $\mathsf{F} : \mathbf{A} \to \mathbf{B}$ that is part of an equivalence of categories must be *full* and *faithful*, i.e. there must be a bijection between the hom-sets $\mathbf{A}(A, A')$ (the set of all morphisms from $A$ to $A'$) and $\mathbf{B}(\mathsf{F}A, \mathsf{F}A')$. Thus our next goal is to define a set of morphisms between each pair of join-semilattices which corresponds bijectively to the set of Scott continuous mappings between the associated algebraic lattices. It is easy to see that we cannot expect to use functions for this purpose for mere cardinality reasons: the set of compacts can be significantly smaller than its algebraic lattice. This problem was already solved by Scott in the closely related case of his *information systems* [Sco82a], which we shall also encounter later on. The idea is to shift to a special set of relations, called *approximable mappings*. To our knowledge, the notion of approximable mappings has not yet been introduced to the study of join-semilattices, so we spell out the details.

**Definition 4.2** Consider join-semilattices $S$ and $T$ with least elements $\bot_S$ and $\bot_T$, respectively. A relation $\rightsquigarrow \,\subseteq S \times T$ is an *approximable mapping* if the following hold:

(am1) $a \rightsquigarrow \bot_T$ (non-emptiness)

(am2) $a \rightsquigarrow b$ and $a \rightsquigarrow b'$ implies $a \rightsquigarrow b \vee b'$ (directedness)

(am3) $a \leq a'$, $a \rightsquigarrow b$, and $b' \leq b$ imply $a' \rightsquigarrow b'$ (monotonicity and downward closure)

for all elements $a, a' \in S$ and $b, b' \in T$. This situation is denoted by writing $S \rightsquigarrow T$.

The labels for the above properties already indicate their purpose: for every element $a \in S$ the set $\{b \in T \mid a \rightsquigarrow b\}$ is an ideal of $T$ and the resulting assignment $S \to \mathsf{Idl}(T)$ is monotone. It is now rather obvious how this encodes Scott continuous functions: The image of a compact element is given explicitly via the ideal of compacts which approximates it. The image of a non-compact element is obtained by representing it as directed supremum of compacts and applying Scott continuity.

Some easy checks show that join-semilattices with least element together with approximable mappings indeed constitute a category $\mathbf{Sem}_\vee$, where composition of morphisms is defined as the usual composition of relations. Thus for two approximable mappings $S \leadsto_1 R$ and $R \leadsto_2 T$, one defines

$$\leadsto_2 \circ \leadsto_1 = \{(s,t) \mid \text{there is } r \in R \text{ such that } (s,r) \in \leadsto_1 \text{ and } (r,t) \in \leadsto_2\}.$$

Clearly, $\leadsto_2 \circ \leadsto_1$ satisfies (am1) since $a \leadsto_1 \perp_R$ and $\perp_R \leadsto_2 \perp_T$. Likewise, under the assumptions of (am2), one finds intermediate values $r, r' \in R$ with $a \leadsto_1 r \leadsto_2 b$ and $a \leadsto_1 r' \leadsto_2 b'$. By (am2) $a \leadsto_1 r \vee r'$, and by (am3) $r \vee r' \leadsto_2 b$ and $r \vee r' \leadsto_2 b'$. Hence $a \leadsto_1 r \vee r' \leadsto_2 b \vee b'$ by another application of (am2). Finally, suppose the assumptions for (am3) hold for $\leadsto_2 \circ \leadsto_1$. Then there is $r \in R$ such that $a \leadsto_1 r \leadsto_2 b$ and hence $a' \leadsto_1 r \leadsto_2 b'$ as required. The identity morphism on a semilattice $S \in |\mathbf{Sem}_\vee|$ is just its greater-or-equal relation $\geq_S$. The fact that this yields an identity under relational composition is just statement (am3). Associativity is inherited from relational composition.

**Lemma 4.3** The object mappings $\mathsf{Idl}$ and $\mathsf{K}$ from Section 3 can be extended to morphisms as follows. For any approximable mapping $\leadsto \subseteq S \times T$, define $\mathsf{Idl}(\leadsto) : \mathsf{Idl}(S) \to \mathsf{Idl}(T)$ as $\mathsf{Idl}(\leadsto)(I) = \{b \mid \text{there is } a \in I \text{ with } a \leadsto b\}$. For any Scott continuous mapping $f : L \to M$, define $\mathsf{K}f \subseteq \mathsf{K}L \times \mathsf{K}M$ by setting $\mathsf{K}f = \{(a,b) \mid b \leq f(a)\}$. These definitions produce functors $\mathsf{Idl} : \mathbf{Sem}_\vee \to \mathbf{Alg}$ and $\mathsf{K} : \mathbf{Alg} \to \mathbf{Sem}_\vee$.

**Proof.** To see that $\mathsf{Idl}$ is indeed well-defined, observe that for any $a \in S$, $\mathsf{Idl}(\leadsto)(\downarrow a) = \{b \mid a \leadsto b\}$, by (am3). This set has already be recognized as an ideal, and hence $\mathsf{Idl}(\leadsto)$ is well-defined for the compact elements of $\mathsf{Idl}(S)$. By algebraicity, any ideal $I$ is equal to the directed union $\bigcup_{a \in I} \downarrow a$, and hence, observing that $\mathsf{Idl}(\leadsto)$ preserves all unions, $\mathsf{Idl}(\leadsto)(I) = \bigcup_{a \in I} \mathsf{Idl}(\leadsto)(\downarrow a)$. This observation shows that, as a directed union of ideals, $\mathsf{Idl}(\leadsto)(I)$ is an ideal, and that $\mathsf{Idl}(\leadsto)$ is Scott continuous.

It is immediate that $\mathsf{Idl}(\leadsto)$ maps the identity approximable mapping $\geq$ to the identity function. To see that it also preserves composition, note that Scott continuity allows us to restrict to the case of principal ideals. Thus consider two approximable mappings $S \leadsto_1 R$ and $R \leadsto_2 T$ and some principal ideal $\downarrow a$, $a \in S$. We compute $(\mathsf{Idl}(\leadsto_2) \circ \mathsf{Idl}(\leadsto_1))(\downarrow a) = \mathsf{Idl}(\leadsto_2)\{r \mid a \leadsto_1 r\} = \{b \mid \text{there is } r \text{ with } a \leadsto_1 r \text{ and } r \leadsto_2 b\} = \{b \mid a(\leadsto_2 \circ \leadsto_1)b\} = \mathsf{Idl}(\leadsto_2 \circ \leadsto_1)(\downarrow a)$.

Now clearly $\mathsf{K}f$ has properties (am1) to (am3). For functoriality consider Scott continuous functions $f_1 : L \to M$ and $f_2 : M \to N$. It is easy to see that for $a \in \mathsf{K}L$ and $c \in \mathsf{K}N$, whenever there is $b \in \mathsf{K}M$ with $b \leq f_1(a)$ and $c \leq f_2(b)$, one has $c \leq f_2(f_1(a))$. Since the converse also holds, we find that $\mathsf{K}(f_2 \circ f_1) = \{(a,c) \mid c \leq f_2(f_1(a))\} = \{(a,c) \mid \text{there is } b \in \mathsf{K}M \text{ with } b \leq f_1(a) \text{ and } c \leq f_2(b)\} =$ $Kf_2 \circ Kf_1$. Finally, applying $K$ to the identity function clearly yields the identity approximable mapping. □

We finish this section by showing the expected categorical equivalence:

**Theorem 4.4** *The functors* Idl *and* K *of Section 3 yield an equivalence of the categories* **Alg** *and* **Sem**$_\vee$.

**Proof.** For an algebraic lattice $L$ let $\eta_L : L \to \mathsf{Idl}(\mathsf{K}(L)) : x \mapsto \downarrow x \cap \mathsf{K}(L)$ be the isomorphism as established in Theorem 3.6. Now consider an algebraic lattice $M$ and a Scott continuous function $f : L \to M$. For any element $x \in L$, $\mathsf{Idl}(\mathsf{K}(f))$ maps the ideal $\eta_L(x)$ to the ideal $\{b \mid \text{there is } a \in \mathsf{K}(L) \text{ with } a \leq x \text{ and } b \leq f(a)\}$. Since Scott continuity guarantees that the supremum of all $f(a)$ is $f(x)$, this is just the set $\eta_M(f(x))$ of all compacts below $f(x)$. Consequently, $\mathsf{Idl}(\mathsf{K}(f))(\eta_L(x)) = \eta_M(f(x))$, i.e. $\eta$ is natural.

For a join-semilattice $S$ with least element, we define $\epsilon_S \subseteq S \times \mathsf{K}(\mathsf{Idl}(S))$ by setting $\epsilon_S = \{(a, I) \mid I \subseteq \downarrow a\}$. From Theorem 3.6 we derive that every compact ideal $I$ is of the form $\downarrow b$, hence $\epsilon_S = \{(a, \downarrow b) \mid b \leq a\}$. It should now be obvious that $\epsilon_S$ is an isomorphism whose inverse is given by $\{(\downarrow b, a) \mid a \leq b\}$. For naturality of $\epsilon$, consider some approximable mapping $S \rightsquigarrow T$. We compute $\mathsf{K}(\mathsf{Idl}(\rightsquigarrow)) \circ \epsilon_S = \{(a, \downarrow b) \mid \text{there is } a' \in S \text{ with } a' \leq a \text{ and } (\downarrow a', \downarrow b) \in \mathsf{K}(\mathsf{Idl}(\rightsquigarrow))\}$. Expanding the condition $(\downarrow a', \downarrow b) \in \mathsf{K}(\mathsf{Idl}(\rightsquigarrow))$, we find it equivalent to $\downarrow b \subseteq \mathsf{Idl}(\rightsquigarrow)(\downarrow a')$, which in turn is true iff $\downarrow b \subseteq \{t \mid a' \rightsquigarrow t\}$, exploiting the fact that $\downarrow a'$ is compact. Finally, by (am3) this is equivalent to $a' \rightsquigarrow b$, and we obtain $\mathsf{K}(\mathsf{Idl}(\rightsquigarrow)) \circ \epsilon_S = \{(a, \downarrow b) \mid a \rightsquigarrow b\}$, again by (am3). On the other hand, $\epsilon_T \circ \rightsquigarrow = \{(a, \downarrow b) \mid \text{there is } b' \in T \text{ with } a \rightsquigarrow b' \text{ and } b \leq b'\}$. Using (am3) once more, this evaluates to $\{(a, \downarrow b) \mid a \rightsquigarrow b\}$, which finishes the proof of naturality of $\epsilon$. □

## 5 A cartesian closed category of formal contexts

Formal concept analysis (FCA, [GW99]) is a powerful lattice-based tool for symbolic data analysis. In essence, it is based on the extraction of a lattice — called *formal concept lattice* — from a binary relation called *formal context* consisting of a set of objects, a set of attributes, and an incidence relation. The transformation from a two-dimensional incidence table to a lattice structure is a crucial *paradigm shift* from which FCA derives much of its power and versatility as a modeling tool. The concept lattices obtained this way turn out to be exactly the complete lattices, and the particular way in which they structure and represent knowledge is very appealing and natural from the perspective of many scientific disciplines.

The successful applications of FCA, however, are mainly restricted to finite contexts and finite concept lattices, since infinite complete lattices generally do not lend themselves for practical implementations. Yet, infinite structures are highly relevant for numerous concrete tasks in knowledge representation and reasoning: model theories of logic programs, computation domains in functional programming, and class hierarchies in ontology research are some typical examples. In order to make methods from FCA available in these application areas, we suggest an interpretation of formal contexts based solely on finitely representable knowledge, thereby obtaining a canonical and computationally feasible representation of infinite data-structures. In effect, we establish a systematic connection between formal concept analysis and algebraic lattices, and thus with domain theory [AJ94], as a categorical equivalence, enriching the link between the two areas as outlined in [Zha03a]. This leads to a category of formal contexts that we now show directly to be cartesian closed.

**Definition 5.1** A *formal context* is a structure $\mathbb{P} = (O, A, \models)$, where $O$ and $A$ are sets, and $\models \subseteq O \times A$ is a binary relation. In this case the members of $O$ are called *objects*, the members of $A$ are called *attributes*, and $\models$ is viewed as an *incidence relation* between these two. Accordingly, one says that an object $o$ *has property* $a$ whenever $o \models a$, i.e. $(o, a) \in \models$.

Functions $\alpha_\mathbb{P} : 2^O \to 2^A$ and $\omega_\mathbb{P} : 2^A \to 2^O$ are defined by setting $\alpha_\mathbb{P}(X) = \{a \in A \mid o \models a \text{ for all } o \in X\}$ and $\omega_\mathbb{P}(Y) = \{o \in O \mid o \models a \text{ for all } a \in Y\}$.[1] If the context is clear, we omit the subscript from these maps. We also abbreviate $\alpha \circ \omega$ by $\alpha\omega$ etc. as is customary in category theory.

Intuitively, $\alpha$ yields all attributes common to a set of objects. Conversely, $\omega$ maps a set of attributes to all objects that fall under all of these attributes. It is straightforward to show that $\alpha$ and $\omega$ form an antitone Galois connection between the powerset lattices. This is usually exploited for constructing closure operators $\alpha \circ \omega : 2^A \to 2^A$ and $\omega \circ \alpha : 2^O \to 2^O$. It turns out that the closure systems for both of these are dually isomorphic, the isomorphisms being given by $\alpha$ and $\omega$.

For studying these closure systems, we can therefore focus our attention on the map $\alpha \circ \omega$. Sets of attributes that are closed with respect to this operator are called (attribute) concepts in the literature. FCA builds on the fact that the collection of all concepts of any given formal context is a complete lattice, and that all complete lattices can be obtained this way. This relationship is mediated by the closure system on $2^A$ induced by the mapping $\alpha\omega$. We take a slightly different approach and focus our attention on the operation of $\alpha\omega$ on $\mathsf{K}(2^A)$, the join-semilattice with least element given by the finite subsets of $A$. It turns out that this way we obtain all complete algebraic lattices instead of all complete ones. Now Corollary 3.10 suggests the following.

---

[1] In FCA, $\alpha_\mathbb{P}(X)$ is usually written as $X'$, and $\omega_\mathbb{P}(Y)$ is similarly written as $Y'$. We feel that for our treatment a more explicit notation is more convenient.

**Corollary 5.2** For every formal context $\mathbb{P} = (O, A, \models)$, the set $\mathsf{Sem}(\mathbb{P}) = \alpha\omega(\mathsf{K}(2^A))$ is a join-semilattice with least element. Conversely, every such semilattice can (up to isomorphism) be represented in this way.

**Proof.** In spite of our earlier considerations, we give the easy direct proof. For two finite sets $X$ and $Y$, $\alpha\omega(X \cup Y)$ is the least closed set that contains $X$ and $Y$, and thus also $\alpha\omega(X)$ and $\alpha\omega(Y)$. Hence $\alpha\omega(X) \vee \alpha\omega(Y) = \alpha\omega(X \cup Y)$. The first part of the proof is finished by noting that $\alpha\omega(\emptyset)$ is the least closed set.

Conversely, for a join-semilattice with least element $S$, consider the context $(S, S, \geq)$. Then for any finite $X \subseteq S$, $\alpha\omega(X)$ is the set of all lower bounds of all upper bounds of $X$. But this is easily recognized as $\downarrow \bigvee X$. Note that the least upper bound of the empty set is just the least element. The obvious isomorphism between $S$ and the semilattice $(\{\downarrow s \mid s \in S\}, \subseteq)$ suffices to complete the proof. □

By Theorem 3.6 the above shows that every algebraic lattice can be represented by some formal context and vice versa. To make this explicit, we can extend the closure operator of Corollary 5.2 to a Scott continuous closure operator on $2^A$, as done before in the proof of Corollary 3.10. In this way we can recover the following result from [ZS0x].

**Corollary 5.3** Consider a formal context $\mathbb{P} = (O, A, \models)$ and the mapping $c : 2^A \to 2^A : x \mapsto \bigcup\{\alpha\omega(X) \mid X \subseteq x, X \text{ finite}\}$. Then $\mathsf{Alg}(\mathbb{P}) = c(2^A)$ is an algebraic lattice and every algebraic lattice is of this form (up to isomorphism).

**Proof.** Clearly, $c$ is just the unique Scott continuous closure operator induced by $\alpha \circ \omega$ as in Corollary 3.10. By Theorem 3.9 its closure system is indeed an algebraic lattice. For the other direction combine Theorem 3.9 and Theorem 3.6 to see that $c(2^A)$ is isomorphic to the ideal completion of $\mathsf{Sem}(\mathbb{P})$. Since every algebraic lattice is of this form for some join-semilattice with least element, the claim follows from Corollary 5.2. □

Closed sets with respect to the operator $c$ from the above proposition have been termed *approximable concepts* in [ZS0x]. Naturally, it is also possible to extend this result to a categorical equivalence. For this purpose we define a category **Cxt** of formal contexts. The morphisms between two contexts $\mathbb{P}$ and $\mathbb{Q}$ are defined by setting $\mathbf{Cxt}(\mathbb{P}, \mathbb{Q}) = \mathbf{Sem}_\vee(\mathsf{Sem}(\mathbb{P}), \mathsf{Sem}(\mathbb{Q}))$.[2] The following is readily seen.

---

[2]In [HZ04] a slightly different definition of morphisms is given. In the formulation given there, the corresponding approximable mapping is not defined on the closed sets $\mathsf{Sem}(\mathbb{P})$ but on all finite attribute sets. We get a context morphism in this sense by extending our approximable mappings, relating two finite sets iff their closures are related.

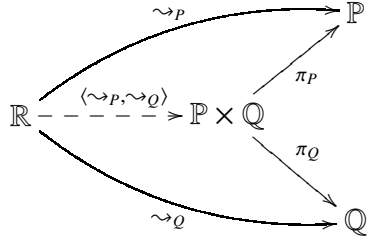

Figure 1: The product construction in **Cxt**.

**Theorem 5.4** The categories $\mathbf{Sem}_\vee$ and **Cxt** are equivalent.

The functors needed for this result are obvious: on the object level, we obtain suitable mapping from Corollary 5.2, and the situation for morphisms is trivial. The construction of the natural isomorphisms is similar to the one of $\epsilon$ in Theorem 4.4, where the identity approximable mapping was modified using the given order-isomorphism of the semilattices.

In the remainder of this section we investigate the categorical constructions that are possible within the categories **Alg**, $\mathbf{Sem}_\vee$, and **Cxt**, where the latter will be the explicit object of study. Because **Cxt** is equivalent to **Alg**, we know that it is *cartesian closed*. We make the required constructions explicit in the sequel, and thus give a mostly self-contained proof of cartesian closedness of **Cxt**.

**Definition 5.5** A category **C** is *cartesian closed* if it has all finite products, and there is a functor $\mathbf{C}^{\mathrm{op}} \times \mathbf{C} \to \mathbf{C} : (A, B) \mapsto B^A$ and a natural bijection between the hom-sets $\mathbf{C}(A \times B, C)$ and $\mathbf{C}(A, C^B)$.

Exact requirements for showing each of these properties will be given in the respective proofs and statements. We first consider the empty product, i.e. the terminal object, which turns out to be given by the formal context $\mathbf{1} = (\emptyset, \emptyset, \emptyset)$. Indeed, for every formal context $\mathbb{P} = (O, A, \models)$ there is a unique approximable mapping $\mathbb{P} \rightsquigarrow \mathbf{1}$ that relates every finite subset of $A$ to the empty set. The situation for binary products is not much more difficult.

**Proposition 5.6** Consider two formal contexts $\mathbb{P} = (O_P, A_P, \models_P)$ and $\mathbb{Q} = (O_Q, A_Q, \models_Q)$, and define a formal context $\mathbb{P} \times \mathbb{Q} = (O_P \uplus O_Q, A_P \uplus A_Q, (\models_Q) \uplus (\models_P) \uplus (O_P \times A_Q) \uplus (O_Q \times A_P))$, where $\uplus$ denotes disjoint union.

Then $\mathbb{P} \times \mathbb{Q}$ is the categorical product of $\mathbb{P}$ and $\mathbb{Q}$, i.e. there are approximable mappings $\pi_P : \mathbb{P} \times \mathbb{Q} \to \mathbb{P}$ and $\pi_Q : \mathbb{P} \times \mathbb{Q} \to \mathbb{Q}$ such that, given approximable mappings $\rightsquigarrow_P$ and $\rightsquigarrow_Q$ as in Figure 1, there is a unique approximable mapping $\langle \rightsquigarrow_P, \rightsquigarrow_Q \rangle$ that makes this diagram commute.

**Proof.** Since context morphisms were defined with reference to the induced semi-lattices, we first look at $\mathsf{Sem}(\mathbb{P} \times \mathbb{Q})$. It is easy to see that concept closure in $\mathbb{P} \times \mathbb{Q}$ is computed by taking disjoint unions of closures in $\mathbb{P}$ and $\mathbb{Q}$, i.e. for sets $X \subseteq A_P$ and $Y \subseteq A_Q$, one finds that $\alpha\omega(X \uplus Y) = \alpha\omega(X) \uplus \alpha\omega(Y)$. Hence every element of $\mathsf{Sem}(\mathbb{P} \times \mathbb{Q})$ corresponds to a unique disjoint union of elements of $\mathsf{Sem}(\mathbb{P})$ and $\mathsf{Sem}(\mathbb{Q})$.

We can now define the projections by setting $(X \uplus Y, X') \in \pi_P$ iff $X' \subseteq X$ and $(X \uplus Y, Y') \in \pi_Q$ iff $Y' \subseteq Y$, for all $X, X' \in \mathsf{Sem}(\mathbb{P})$ and $Y, Y' \in \mathsf{Sem}(\mathbb{Q})$. It is readily seen that these morphisms satisfy the properties of Definition 4.2.

Now consider $\leadsto_P$ and $\leadsto_Q$ as in Figure 1. We define the relation $\langle \leadsto_P, \leadsto_Q \rangle$ by setting $(Z, X \uplus Y) \in \langle \leadsto_P, \leadsto_Q \rangle$ iff $Z \leadsto_P X$ and $Z \leadsto_Q Y$, for all concepts $X$, $Y$, $Z$ from the corresponding semilattices. Again it is easy to check the conditions of Definition 4.2, since they follow immediately from the corresponding properties of $\leadsto_P$ and $\leadsto_Q$. Furthermore, if there is $X \uplus Y \in \mathsf{Sem}(\mathbb{P} \times \mathbb{Q})$ with $(Z, X \uplus Y) \in \langle \leadsto_P, \leadsto_Q \rangle$ and $(X \uplus Y, X') \in \pi_P$ then $Z \leadsto_P X'$ by the definition of $\pi_P$ and (am3). Conversely, if $Z \leadsto_P X'$ then one finds that $X' \uplus \alpha\omega(\emptyset) \in \mathsf{Sem}(\mathbb{P} \times \mathbb{Q})$ yields the required intermediate element to show that $(Z, X') \in \pi_P \circ \langle \leadsto_P, \leadsto_Q \rangle$. Since a similar reasoning applies to $\leadsto_Q$, Figure 1 commutes as required.

Finally, for uniqueness of $\langle \leadsto_P, \leadsto_Q \rangle$ consider $\mathbb{R} \leadsto \mathbb{P} \times \mathbb{Q}$ with $\pi_P \circ \leadsto = \leadsto_P$ and $\pi_Q \circ \leadsto = \leadsto_Q$. If $Z \leadsto X \uplus Y$, then $(Z, X) \in \pi_P \circ \leadsto$ and hence $Z \leadsto_P X$ and, by a similar reasoning, $Z \leadsto_Q Y$. Conversely, if $Z \leadsto_P X$ then there must be some $X'$ and $Y'$ such that $X \subseteq X'$ and $Z \leadsto X' \uplus Y'$. By (am3) this implies $Z \leadsto X \uplus Y'$. The same argument can be applied to $\leadsto_Q$. Thus whenever $Z \leadsto_P X$ and $Z \leadsto_Q Y$, there are $X'$ and $Y'$ with $Z \leadsto X \uplus Y'$ and $Z \leadsto X' \uplus Y$. Invoking properties (am2) and (am3) for $\leadsto$, this shows that $Z \leadsto X \uplus Y$. We have just shown that $Z \leadsto X \uplus Y$ iff $Z \leadsto_P X$ and $Z \leadsto_Q Y$, and hence that $\leadsto = \langle \leadsto_P, \leadsto_Q \rangle$ as required. □

The above product construction is also known in formal concept analysis as the *direct sum* of two contexts [GW99]. However, it is not the only possible specification of the products in **Alg**. For each formal context $\mathbb{P} = (O_P, A_P, \models_P)$, we define a context $\mathbb{P}^+ = (O_P^+, A_P^+, \models_{P^+})$, where $O_P^+ = O_P \cup \{g\}$ and $A_P^+ = A_P \cup \{m\}$, with $g$ and $m$ being fresh elements: $g \notin O_P$ and $m \notin A_P$. For defining the incidence relation, we set $o \models_{P^+} a$ whenever $o \models_P a$ (requiring that $a \in A_P$ and $o \in O_P$) or $o = g$ or $a = m$. Thus $\mathbb{P}^+$ emerges from $\mathbb{P}$ by "adding a full row and a full column."

Now let $\mathbb{P} = (O_P, A_P, \models_P)$ and $\mathbb{Q} = (O_Q, A_Q, \models_Q)$ be formal contexts. Define a new formal context $\mathbb{P} \otimes \mathbb{Q} = (O_P^+ \times O_Q^+, A_P^+ \times A_Q^+, \models_{P \times Q})$ of $\mathbb{P}$ and $\mathbb{Q}$ by setting $(o_1, o_2) \models_{P \times Q} (a_1, a_2)$ iff $o_1 \models_{P^+} a_1$ and $o_2 \models_{Q^+} a_2$. This turns out to be an alternative description of the products in **Cxt**.

**Proposition 5.7** Given arbitrary formal contexts $\mathbb{P} = (O_P, A_P, \models_P)$ and $\mathbb{Q} = (O_Q, A_Q, \models_Q)$, the contexts $\mathbb{P} \times \mathbb{Q}$ and $\mathbb{P} \otimes \mathbb{Q}$ are isomorphic in **Cxt**. Equivalently, $\mathbb{P} \otimes \mathbb{Q}$ is the object part of the categorical product of $\mathbb{P}$ and $\mathbb{Q}$ in **Cxt**.

**Proof.** The required isomorphism corresponds to an iso approximable mapping between the semilattices $\mathsf{Sem}(\mathbb{P} \times \mathbb{Q})$ and $\mathsf{Sem}(\mathbb{P} \otimes \mathbb{Q})$. The elements of the former were already recognized as disjoint unions of concepts from $\mathbb{P}$ and $\mathbb{Q}$. In the latter case, concepts are easily recognized as products of concepts from $\mathbb{P}^+$ and $\mathbb{Q}^+$. Adding the additional elements $m$ and $g$ guarantees that neither of these extended formal contexts allows for the empty set as a concept, so that each element of $\mathsf{Sem}(\mathbb{P} \otimes \mathbb{Q})$ is indeed of the form $X \times Y$ for two uniquely determined concepts $X = \alpha\omega(X) \in \mathsf{Sem}(\mathbb{P}^+)$ and $Y = \alpha\omega(Y) \in \mathsf{Sem}(\mathbb{Q}^+)$.

We define a relation $\leadsto^+ \subseteq \mathsf{Sem}(\mathbb{P} \times \mathbb{Q}) \times \mathsf{Sem}(\mathbb{P} \otimes \mathbb{Q})$ by setting $X \leadsto^+ Y$ whenever $p_1(Y) \cap A_P \subseteq X$ and $p_2(Y) \cap A_Q \subseteq X$, where $p_i$ denotes the projection to the $i$th components in a set of pairs. Conversely, a relation $\leadsto^- \subseteq \mathsf{Sem}(\mathbb{P} \otimes \mathbb{Q}) \times \mathsf{Sem}(\mathbb{P} \times \mathbb{Q})$ is specified by setting $Y \leadsto^- X$ whenever $X \cap A_P \subseteq p_1(Y)$ and $X \cap A_Q \subseteq p_2(Y)$.

We claim that $\leadsto^+$ and $\leadsto^-$ are mutually inverse approximable mappings between $\mathsf{Sem}(\mathbb{P} \times \mathbb{Q})$ and $\mathsf{Sem}(\mathbb{P} \otimes \mathbb{Q})$. The properties of Definition 4.2 follow immediately from our use of set-theoretic operations in the definitions. Furthermore it is easy to see that $X(\leadsto^- \circ \leadsto^+)X'$ implies $X' \subseteq X$ for any two elements $X, X' \in \mathsf{Sem}(\mathbb{P} \times \mathbb{Q})$. The converse implication also holds, which can be concluded from the obvious relationships $X \leadsto^+ \alpha\omega(X \cap A_P) \times \alpha\omega(X \cap A_Q)$, $\alpha\omega(X' \cap A_P) \times \alpha\omega(X' \cap A_Q) \leadsto^- X'$, and $\alpha\omega(X' \cap A_P) \times \alpha\omega(X' \cap A_Q) \subseteq \alpha\omega(X \cap A_P) \times \alpha\omega(X \cap A_Q)$. Hence $\leadsto^- \circ \leadsto^+$ is indeed the identity approximable mapping. A similar reasoning shows that the same is true for $\leadsto^+ \circ \leadsto^-$, thus finishing the proof.

Finally, the assertion that this makes $\otimes$ an alternative product construction is a basic fact from category theory. The required projections are obtained by composing $\leadsto^-$ with the projections from the proof of Proposition 5.6. □

The construction of exponentials in **Cxt** turns out to be slightly more intricate. To fully understand the following definition, it is helpful to look at the function spaces in **Alg**. These are just the sets of all Scott continuous maps between two algebraic lattices under the pointwise order of functions. The standard technique for describing the compact elements of this lattice are so-called *step functions*. Given two algebraic lattices $L$ and $M$ and two compacts $a \in \mathsf{K}(L)$ and $b \in \mathsf{K}(M)$, one defines a function $|a \Rightarrow b| : L \to M$, that maps an element $x$ to $b$ whenever $a \leq x$, and to $\bot_M$ otherwise. It is well-known that any such step function is Scott continuous and compact in the function space of $L$ and $M$ (see [GHK+03]). However, not all compacts are of this form, since finite joins of step functions are also compact maps that can usually take more than two different values.

Our goal is to construct a formal context that represents the join-semilattice of all compact Scott continuous functions in the sense of Corollary 5.2. Intuitively, the collection of all step functions suggests itself as the set of attributes. Finitely generated concepts should represent finite joins of step functions, which in turn correspond bijectively to lower sets with respect to the pointwise order of step functions. In order to obtain a formal context that yields this lower closure, one is tempted to take some subset of Scott continuous functions for objects, and to employ the inverted pointwise order as an entailment relation. This is indeed feasible, but our supply of step functions unfortunately is insufficient to serve as object set in this case. We end up with the following definition:

**Definition 5.8** Consider two formal contexts $\mathbb{P}$ and $\mathbb{Q}$, and the sets $A = \mathsf{Sem}(\mathbb{P}) \times \mathsf{Sem}(\mathbb{Q})$ and $O = \mathsf{Fin}(A)$. A formal context $[\mathbb{P} \rightsquigarrow \mathbb{Q}] = (O, A, \models)$ is defined by setting $\{(a_i, b_i)\} \models (a, b)$ iff $b \subseteq \bigvee \{b_i \mid a_i \subseteq a\}$, where $\bigvee$ is the join operation from the semilattice $\mathsf{Sem}(\mathbb{Q})$.

This definition derives from the above discussion by representing step functions $|a \Rightarrow b|$ via pairs $(a, b)$.[3] Hence, the approximable concepts of $[\mathbb{P} \rightsquigarrow \mathbb{Q}]$ as obtained in Corollary 5.3 are sets of such pairs, i.e. relations between $\mathsf{Sem}(\mathbb{P})$ and $\mathsf{Sem}(\mathbb{Q})$. The reader's suspicion about the true nature of these relations shall be confirmed:

**Lemma 5.9** Given contexts $\mathbb{P}$ and $\mathbb{Q}$, the algebraic lattice $L = \mathsf{Alg}[\mathbb{P} \rightsquigarrow \mathbb{Q}]$ of approximable concepts of $[\mathbb{P} \rightsquigarrow \mathbb{Q}]$ coincides with the lattice of all approximable mappings from $\mathbb{P}$ to $\mathbb{Q}$, ordered by subset inclusion.

**Proof.** Consider any approximable concept $x \in L$. Definition 5.8 implies that the pairs of arbitrary elements $a \in \mathsf{Sem}(\mathbb{P})$ and the least element of $\mathsf{Sem}(\mathbb{Q})$ are modelled by any object of $[\mathbb{P} \rightsquigarrow \mathbb{Q}]$, i.e. (am1) of Definition 4.2 holds for $x$. For (am2), assume $(a, b_1) \in x$ and $(a, b_2) \in x$. Following the construction in Corollary 5.3, one finds that $\alpha\omega(\{(a, b_1), (a, b_2)\}) \subseteq x$. However, for any object $o$ of $[\mathbb{P} \rightsquigarrow \mathbb{Q}]$, $o \models (a, b_1)$ and $o \models (a, b_2)$ clearly implies $o \models (a, b_1 \vee b_2)$, by expanding the definition of $\models$, and thus $(a, b_1 \vee b_2) \in x$. Finally, for (am3) consider some $(a, b) \in x$, $a' \supseteq a$, and $b' \subseteq b$. Clearly, we have $\alpha\omega(\{(a, b)\}) \subseteq x$. The definition of $\models$ shows immediately that every object that models $(a, b)$ must also model $(a', b')$, and thus $(a', b') \in \alpha\omega(\{(a, b)\})$ as required.

For the converse consider any approximable mapping $\mathbb{P} \rightsquigarrow \mathbb{Q}$. We show that $\rightsquigarrow \in L$. Given any finite subset $X = \{(a_i, b_i)\} \subseteq \rightsquigarrow$, one finds that $X \models (a_n, b_n)$ for all $(a_n, b_n) \in X$. Thus $X \in \omega(X)$ and, whenever $(a, b) \in \alpha\omega(X)$, one also has $X \models (a, b)$, i.e. $b \subseteq \bigvee \{b_j \mid a_j \subseteq a\}$. Defining $J = \{j \mid a_j \subseteq a\}$, one finds that for every $n \in J$, $a_n \subseteq \bigvee \{a_j \mid j \in J\}$ and hence $\bigvee \{a_j \mid j \in J\} \rightsquigarrow b_n$ by (am3). Since $J$ is finite, one can employ an easy induction to show that $\bigvee \{a_j \mid j \in J\} \rightsquigarrow \bigvee \{b_j \mid j \in$ $J\}$, where the case $J = \emptyset$ follows from (am1) and the induction step uses (am2). Obviously $\bigvee\{a_j \mid j \in J\} \subseteq a$ and $b \subseteq \bigvee\{b_j \mid j \in J\}$, and hence $a \rightsquigarrow b$ by (am3). This shows that $\rightsquigarrow$ is an approximable concept. □

---

[3]This correspondence is not injective. In fact, the context $[\mathbb{P} \rightsquigarrow \mathbb{Q}]$ in general contains both duplicate rows and duplicate columns.

The above considerations shed additional light on approximable mappings in general: they can in fact be viewed as lower sets of step functions, the joins of which uniquely determine an arbitrary Scott continuous map between the induced algebraic lattices. We remark that this also hints at an alternative formulation of the constructions in Lemma 4.3.

It remains to show that the above construction does indeed yield a function space in the sense of category theory:

**Proposition 5.10** The construction $[- \rightsquigarrow -]$ yields the categorical function space of the two contexts, i.e. for all contexts $\mathbb{P}$, $\mathbb{Q}$, and $\mathbb{R}$, there is a bijection between the sets $\mathbf{Cxt}(\mathbb{P} \times \mathbb{Q}, \mathbb{R})$ and $\mathbf{Cxt}(\mathbb{P}, [\mathbb{Q} \rightsquigarrow \mathbb{R}])$, and this bijection is natural in all arguments.

**Proof.** Our earlier results can be employed to simplify this proof. The algebraic lattices associated with the above contexts is denoted by $L = \mathsf{Alg}(\mathbb{P})$, $M = \mathsf{Alg}(\mathbb{Q})$, and $N = \mathsf{Alg}(\mathbb{R})$, and we write $[M \to N]$ for the lattice of all Scott continuous functions from $M$ to $N$, ordered pointwise. The categorical equivalences between $\mathbf{Cxt}$, $\mathbf{Sem}_\vee$, and $\mathbf{Alg}$ (Theorem 4.4 and Theorem 5.4) and the categorical role of the product construction $\mathbb{Q} \times \mathbb{R}$ (Proposition 5.6) establish natural bijections between the sets $\mathbf{Cxt}(\mathbb{P} \times \mathbb{Q}, \mathbb{R})$ and $\mathbf{Alg}(L \times M, N)$, where $L \times N$ is the standard product order. Likewise, using the same equivalences and the bijection of function spaces from Lemma 5.9, one finds another natural bijection between $\mathbf{Cxt}(\mathbb{P}, [\mathbb{Q} \rightsquigarrow \mathbb{R}])$ and $\mathbf{Alg}(L, [N \to M])$.

The proof is completed by providing the well-known natural bijection of the sets $\mathbf{Alg}(L \times M, N)$ and $\mathbf{Alg}(L, [N \to M])$. This standard proof can for example be found in [GHK+03]. □

Summing up these results, we obtain:

**Theorem 5.11** The categories $\mathbf{Alg}$, $\mathbf{Sem}_\vee$, and $\mathbf{Cxt}$ are cartesian closed.

**Proof.** $\mathbf{Cxt}$ was shown cartesian closed in Proposition 5.6 and Proposition 5.10. Closure of the other categories follows by their categorical equivalence (Theorem 4.4 and Theorem 5.4). □

The cartesian closed category $\mathbf{Cxt}$ which we propose here is tailored to the needs of Computer Science. It differs from the categories normally considered in formal concept analysis by emphasizing *algebraicity*, whereas morphisms listed e.g. in [GW99] are suitable for complete, but not necessarily algebraic, lattices.

We also stress the fact that our novel interpretation of formal contexts perfectly agrees with the classical one, as long as finite contexts or lattices are considered, which covers most of the current FCA applications in Computer Science. On the other hand, the different treatment of infinite data structures displays a deviation from the classical philosophically motivated viewpoint towards one that respects the practical constraints of finiteness and computability.

# 6 Further representations

So far, we encountered three equivalent representations for algebraic lattices. Clearly, the hard part was to establish the equivalence of the rather diverse categories **Alg** and **Sem**$_\vee$. Many other equivalent categories can now be recognized by relating them to one of these two — an objective that will in general be accomplished rather easily. A typical example for this has already been given in form of the category **Cxt**, that was easily seen to be equivalent to **Sem**$_\vee$.

The representations given below are grouped according to these observations: we start with "logical" descriptions that have their closest relationships to the categories **Cxt** and **Sem**$_\vee$, and then proceed to formulations that can be connected to **Alg** in a more natural way. Classifying representations in this way is by no means arbitrary: as we will see the end of this section, our arrangement reflects the "localic" respectively "spacial" side of a very specific case of Stone duality.

## 6.1 Logic and information systems

The representation of join-semilattices via formal contexts did already incorporate some logical flavor: approximable concepts can be viewed as sets closed under a certain entailment relation. Scott continuity of this closure is reminiscent of the compactness property of a logic. However, we will see that a much closer connection to some very well-known logics can be made. The reader is referred to [DH01] for related considerations.

**Definition 6.1** Given a set $A$ of propositions, the set of well-formed *conjunctive propositional formulae* $\mathscr{S}(A)$ over $A$ is given by the following expression:

$$\mathscr{S}(A) ::= \top \mid a \in A \mid (\mathscr{S}(A) \wedge \mathscr{S}(A))$$

A relation $\vdash \,\subseteq \mathscr{S}(A) \times \mathscr{S}(A)$ is a *consequence relation* of conjunctive propositional logic (**CCP** logic) if it is closed under application of the following rules:

$$F \vdash \top \quad \text{(T)} \qquad F \vdash F \quad \text{(R)} \qquad \frac{F \vdash G, \ G \vdash H}{F \vdash H} \quad \text{(Cut)}$$

$$\frac{F \vdash (G \wedge H)}{F \vdash G} \quad \text{(W1)} \qquad \frac{F \vdash (G \wedge H)}{F \vdash H} \quad \text{(W2)} \qquad \frac{F \vdash G,\ F \vdash H}{F \vdash (G \wedge H)} \quad \text{(And)}$$

In this case $(\mathscr{S}(A), \vdash)$ is called a *deductive system* (of CCP logic). For any two formulae $F, G \in \mathscr{S}(A)$, the situation where $F \vdash G$ and $G \vdash F$ is denoted $F \approx G$.

Hence deductive systems are logical systems of the conjunctive fragment of propositional logic, together with a (not necessarily minimal) consequence relation. The following properties are easily verified.

**Lemma 6.2** Consider a deductive system $(\mathscr{S}(A), \vdash)$. The following hold for all formulae $F$, $G$, and $H \in \mathscr{S}(A)$:

- $((F \wedge G) \wedge H) \approx (F \wedge (G \wedge H))$

- $(F \wedge G) \approx (G \wedge F)$

- $F \approx (F \wedge F)$

- $F \approx (F \wedge \top)$

Hence we see that the rules (W1), (W2), and (And) imply associativity, commutativity, and idempotency of $\wedge$. Furthermore, occurrences of $\top$ can be eliminated. Consequently, we henceforth write formulae of CCP in the form $a_1 \wedge a_2 \wedge \ldots \wedge a_n$ ($a_i \in A$), knowing that this determines a set of "real" formulae up to proof-theoretic equivalence. Additionally, for the case $n = 0$ the above expression is interpreted as the singleton set $\{\top\}$. Any statement about formulae in this notation represents the corresponding set of statements about the original formulae. We can now consider the algebraic semantics (see [DH01]) of these logics. This is based largely on the following notion:

**Definition 6.3** Consider a deductive system $(\mathscr{S}(A), \vdash)$. The *Lindenbaum algebra* of $(\mathscr{S}(A), \vdash)$ is the poset obtained from the preorder $(\mathscr{S}(A), \vdash)$ through factorization by the equivalence relation $\approx$, i.e. $[F]_\approx \leq [G]_\approx$ iff $F \vdash G$. The Lindenbaum algebra is denoted by $\mathsf{LA}(\mathscr{S}(A), \vdash)$.

Hence the Lindenbaum algebra is a partially ordered set of $\approx$-equivalence classes of formulae, ordered by syntactic entailment. Since it can cause hardly any confusion, we take the freedom to denote equivalence classes by one of their representatives or even by the simplified notation introduced above. Of course, this creates possible ambiguity between the conjunction symbol and the meet operation within the Lindenbaum algebra. The following lemma shows that this is not a problem.

**Lemma 6.4** Consider a deductive system $(\mathscr{S}(A), \vdash)$ and formulae $F, G \in \mathscr{S}(A)$. Then $[F]_\approx \wedge [G]_\approx = [F \wedge G]_\approx$.

**Proof.** We have to show that $F \wedge G \vdash F$, $F \wedge G \vdash G$, and that for any formula $H$ such that $H \vdash F$ and $H \vdash G$, we find $H \vdash F \wedge G$. These assertions are obvious consequences of the proof rules of CCP. □

Since the meet operation yields a unique result, this shows that $F \approx F'$ and $G \approx G'$ imply $F \wedge G \approx F' \wedge G'$, which is just the *Replacement Theorem* [DH01] for CCP logics. We state the now obvious representation theorem:

**Theorem 6.5** *For any deductive system $(\mathscr{S}(A), \vdash)$, the Lindenbaum algebra $\mathsf{LA}(\mathscr{S}(A), \vdash)$ is a meet-semilattice with greatest element. Conversely, every such semilattice is isomorphic to the Lindenbaum algebra of some deductive system.*

**Proof.** Lemma 6.4 already showed the existence of binary meets. We conclude the first part of the proof by noting that $[\top]_\approx$ is the required greatest element.

For the converse let $S$ be a meet-semilattice with greatest element. We define a consequence relation $\vdash$ on $\mathscr{S}(S)$ by setting, for all $a_1, a_2, \ldots, a_n, b_1, b_2, \ldots, b_m \in S$, $a_1 \wedge a_2 \wedge \ldots \wedge a_n \vdash b_1 \wedge b_2 \wedge \ldots \wedge b_m$ whenever $a_1 \wedge a_2 \wedge \ldots \wedge a_n \leq b_1 \wedge b_2 \wedge \ldots \wedge b_m$. One can easily check that this definition satisfies all of the required rules. Note that (T) follows by our convention to represent $\top$ by the empty conjunction. To reduce confusion, we denote meets in $S$ by $\bigwedge$ and meets in $\mathsf{LA}(\mathscr{S}(S), \vdash)$ by $\bigwedge_\approx$.

We claim that $S$ is isomorphic to $\mathsf{LA}(\mathscr{S}(S), \vdash)$. Indeed, one can define mappings $f: S \to \mathsf{LA}(\mathscr{S}(S), \vdash)$ and $g: \mathsf{LA}(\mathscr{S}(S), \vdash) \to S$ by setting $f(a) = [a]_\approx$ and, for propositions $a_i, 1 \leq i \leq n$, $g[\bigwedge_\approx a_i]_\approx = \bigwedge a_i$. To see that $g$ is well-defined, note that for any two formulae $\bigwedge_\approx a_i, \bigwedge_\approx b_j \in \mathscr{S}(S)$ we have that $\bigwedge_\approx a_i \approx \bigwedge_\approx b_j$ (in $\mathscr{S}(S)$) implies $\bigwedge a_i = \bigwedge b_j$ (in $S$) by the definition of $\vdash$.

Finally, we show that $g$ and $f$ are inverse to each other. By what was said above, $g(f(a)) = a$ is immediate. On the other hand, any formula $\bigwedge_\approx a_i$ is syntactically equivalent to $\bigwedge a_i$ by the definition of $\vdash$. This shows bijectivity of $f$ and $g$. Monotonicity of both functions is obvious from their definition. □

This relationship closes the gap to our prior category $\mathbf{Sem}_\vee$, since the above meet-semilattices are just the order duals of the objects within this category. By an approximable mapping between two meet-semilattices with least element or two deductive systems of CCP logic, we mean an approximable mapping between the induced join-semilattices. The following is immediate.

**Theorem 6.6** *Consider the categories $\mathbf{Sem}_\wedge$ and $\mathbf{CCP}$ of meet-semilattices with greatest element and deductive systems of CCP logic, respectively, together with approximable mappings as morphisms. Then $\mathbf{Sem}_\vee$, $\mathbf{Sem}_\wedge$, and $\mathbf{CCP}$ are equivalent.*

The insights just obtained allow to relate our study with results obtained in [HW03, Hit04], where the conjunctive fragment of the logic RZ (introduced in [RZ01]), was found to be closely related to concept closure in FCA. We derive a very similar result, but some preparations are needed first.

An *algebraic cpo D* is a dcpo with least element $\bot$ such that every $e \in D$ is the directed supremum of all compact elements below it. A *coherent algebraic cpo* is an algebraic cpo such that, with respect to the Scott topology (see Definition 6.10), the intersection of any two compact open sets is compact open.

These notions can be found in [RZ01], along with a characterization of the Smyth Powerdomain of any given coherent algebraic cpo $D$ by means of a logic defined on $D$, which we call the *logic RZ*. We will only be concerned with the conjunctive fragment of RZ, which can be given as follows. For compact elements $c_1, \ldots, c_n, d_1, \ldots, d_m$ we write $c_1 \wedge \ldots \wedge c_n \vdash d_1 \wedge \ldots \wedge d_m$ iff any minimal upper bound of $\{c_1, \ldots, c_n\}$ is above all $d_i$. This way, we obtain a deductive system $(\mathsf{K}(D), \vdash)$, and the following result, which is related to those in [HW03, Hit04], and such considerations were put to use in [Hit04] for developing a generic non-monotonic rule-based reasoning paradigm over hierarchical knowledge.

**Theorem 6.7** Let $\mathbb{P} = (O, A, \models)$ be any formal context. Then there is a coherent algebraic cpo $D$ and a mapping $\iota : A \to D$ such that for every finite set $X = \{a_1, \ldots, a_n\} \subseteq A$ we have $\alpha\omega(X) = \{a \mid \iota(a_1) \wedge \ldots \wedge \iota(a_n) \vdash \iota(a)\}$.

**Proof.** Define $D = \mathsf{Alg}(\mathbb{P})$ and set $\iota(a) = \alpha\omega(\{a\})$ for $a \in A$. Since $D$ is a complete algebraic lattice, it is a coherent algebraic cpo.

Now consider the finite set $X$ as above. Using the completeness of the lattice, we obtain that $\iota(X)$ has $\alpha\omega(X)$ as supremum, which suffices. □

The difference between Theorem 6.7 and the results in [HW03, Hit04] lies in the fact that the latter were proven by taking $D$ to be a sublattice of the (classical) formal concept lattice, instead of $\mathsf{Alg}(\mathbb{P})$, which facilitates reasoning with formal contexts in a natural way.

Finally, we come to another popular description of algebraic lattices, that fits well into the above discussion, and will also shed additional light on morphisms of **CCP**.

**Definition 6.8** Consider a structure $(A, \Vdash)$, where $A$ is a set, and $\Vdash \subseteq \mathsf{Fin}(A) \times A$ is a relation between finite subsets of $A$ and elements of $A$. Then $(A, \Vdash)$ is a *Scott information system* (with trivial consistency predicate) if the following hold:

(ISi) $a \in X$ implies $X \Vdash a$,

(ISii) if $X \Vdash y$ for all $y \in Y$ and $Y \vdash a$, then $X \Vdash a$.

Scott information systems were introduced in [Sco82a] as a logical characterization of order structures arising in denotational semantics. The connection to CCP logic is as follows.

**Proposition 6.9** There is a bijective relationship between Scott information systems and deductive systems of CCP logic.

**Proof.** Consider a Scott information system $(A, \Vdash)$. Using the set $A$ as propositions, we obtain the set of CCP formulae $\mathscr{S}(A)$. A consequence relation $\vdash$ for $\mathscr{S}(A)$ is defined by setting $a_1 \wedge a_2 \wedge \ldots \wedge a_n \vdash b_1 \wedge b_2 \wedge \ldots \wedge b_m$ whenever $\{a_1, a_2, \ldots, a_n\} \Vdash b_i$ for all $i = 1, \ldots m$. We have to verify that $\vdash$ is closed under the rules given in Definition 6.1. For the case $m = 0$ the condition is obviously true so that we obtain axiom (T). Likewise, the conditions for axiom (R) are satisfied due to condition (ISi) in Definition 6.8. Similarly, the (Cut) rule follows immediately from (ISii). For the rules (W1), (W2), and (And), we simply notice that these are direct consequences from our definition of $\vdash$.

Now for the opposite direction, consider a deductive system $(\mathscr{S}(A), \vdash)$. Using the set of propositions of $\mathscr{S}(A)$ as attributes, we construct a Scott information system $(A, \Vdash)$, where we define $\{a_1, a_2, \ldots, a_n\} \Vdash b$ whenever $a_1 \wedge a_2 \wedge \ldots \wedge a_n \vdash b$. Again it is straightforward to check that this is indeed an information system. (ISi) can be deduced from the rules (R) and iterated applications of (W1) and (W2). Under the assumption of (ISii), we see that the (And) rule allows us to construct a conjunction that corresponds to the premise $Y$ of the second rule. By (Cut) this yields the required entailment.

To complete the proof, we note that these two constructions are in fact inverse to each other. The identity on Scott information systems is trivial. For CCP logics, we note that any sequent $a_1 \wedge a_2 \wedge \ldots \wedge a_n \vdash b_1 \wedge b_2 \wedge \ldots \wedge b_m$ induces via (W1)/(W2) the existence of sequents $a_1 \wedge a_2 \wedge \ldots \wedge a_n \vdash b_i$, for all $i = 1, \ldots, m$. The original sequent can then be reconstructed from the entailment of the Scott information system induced from these relations. □

Note that this proposition yields a bijective correspondence, not just a relationship up to isomorphism. Indeed Scott information systems are essentially an efficient formulation of conjunctive propositional logic, where the properties of $\wedge$ are obtained implicitly by using sets in the first place. The category of Scott information systems and approximable mappings between the induced semilattices is denoted **SIS**[4]. From 6.9 one easily concludes that **SIS** is *isomorphic* to **CCP**, and hence also equivalent to all categories mentioned earlier.

Furthermore, approximable mappings between CCP logics need not be expressed on the level of their Lindenbaum algebras, but could be formulated directly on formulae. From this viewpoint, approximable mappings appear as consequence relations between different logical languages. Indeed, all the requirements of Definition 4.2 do still have a very intuitive reading under this interpreta-tion: (am1) and (am2) correspond to (T) and (And) of Definition 6.1, respectively, while (am3) can be viewed as a modified (Cut) rule, that also subsumes (W1) and (W2). Hence we recognize approximable mappings as a simple form of *multilingual sequent calculi* as introduced in [JKM99] for the more complicated case of *positive logic* (i.e., logics including conjunction and disjunction). Further details and motivation can be found therein.

---

[4]Historically, this is indeed the first context for which *approximable mappings* were defined [Sco82a].

We remark that one could as well have connected CCP logic or information systems directly to algebraic lattices, instead of presenting the ideal completion for semilattices of compacts. In the case of logics, algebraic lattices are obtained directly as sets of *models* of a deductive system, where models are considered as deductively closed sets of (true) formulae. These turn out to be exactly the filters[5] within the corresponding Lindenbaum algebras, and the duality to ideal completion is immediate. The reader may care to consult [DH01] for a general treatment of such matters. For Scott information systems, algebraic lattices are constructed similarly as sets of *elements*. As defined in [Sco82a], an element of an information system $(A, \Vdash)$ is a subset $x \subseteq A$ such that $a \in x$ whenever there is some finite set $X \subseteq x$ with $X \Vdash a$.

Our logical considerations can also be put to practical use by noting that every *definite logic program* (see, e.g., [Llo87]) can be expressed by a deductive system in the above sense. This has also been mentioned in [Zha03b]. Considering the fact that the theory of definite logic programs is quite well-developed, these insights are merely providing some further explanation for the situation in this field. In the light of the connections to Stone duality outlined below and the immediate connection to algebraic semantics of logical systems, one could also further analyze the situation for more expressive logical languages from this perspective.

Note that only a small portion of Scott information systems and algebraic lattices can be obtained from definite logic programs. The reason is that there are only countably many different programs, but uncountably many Scott information systems (even for countable sets of generators). We also remark that, while algebraicity always makes fixed point computation possible in theory, the specific structure of the information systems of logic programs is employed to ensure that the semantic operator suitable for logic programs is indeed effectively computable.

We do not bother to give a category of logic programs, although this could be done by adjusting the formalism of approximable mappings. However, it is not clear at the moment how the subcategory of algebraic lattices that arises in this way can be characterized.

---

[5]A filter $F$ is the dual of an ideal: an upper set $F = \uparrow F$ such that $a, b \in F$ implies the existence of some $c \in F$ such that $c \leq a$ and $c \leq b$.

## 6.2 The Scott topology

Next we want to study the spacial side of Stone duality. It is here where we find the models and their semantic entailment, while the *localic* side is inhabited by syntactic representations and their proof theory. We already mentioned that models in our case take the specific form of algebraic lattices, and thus it is natural to ask which subsets of models correspond to a logical theory. The appropriate collection of sets turns out to be the following well-known topology:

**Definition 6.10** Consider a dcpo $P$. A subset $O \subseteq P$ is *Scott open* if the following hold:

(i) $x \in O$ and $x \leq y$ imply $y \in O$ ($O$ is an upper set),

(ii) for any directed set $D \subseteq P$, $\bigvee D \in O$ implies $D \cap O \neq \emptyset$ ($O$ is inaccessible by directed suprema).

The *Scott topology* on $P$ is the collection of Scott open sets. We use $\sigma(P)$ to denote this collection and $\Sigma(P) = (P, \sigma(P))$ for the resulting topological space.

But one can also reverse the process to obtain orders from topologies:

**Definition 6.11** Consider a topological space $(X, \tau)$. Then $\tau$ defines a *specialization (pre-)order* $\leq$ on $X$ by setting $x \leq y$ whenever $x \in O$ implies $y \in O$ for any $O \in \tau$. A topology on a partially ordered set is called *order consistent* if its specialization order coincides with the order of the poset.

For an algebraic lattice, the Scott topology has some more specific properties. Recall that an open set is *compact* if it is a compact element of the open set lattice, and that a topology is *coherent* if the intersection of any two compact open sets is compact. Proof for the following statements can be found in [AJ94, GHK$^+$03, Joh82].

**Proposition 6.12** Consider an algebraic lattice $L$. We have the following:

(i) $\Sigma(L)$ is order consistent.

(ii) The set $B = \{\uparrow c \mid c \in \mathsf{K}(L)\}$ is a base for $\sigma(L)$.

(iii) The compact opens of $\Sigma(L)$ are exactly the finite unions of members of $B$.

(iv) $\sigma(L)$ is coherent.

(v) $\sigma(L)$ is sober.[6]

---

[6]We did not define sobriety in this document. Readers who are not familiar with this concept may safely ignore this statement.

Order consistency insures that algebraic lattices and their Scott topologies uniquely characterize each other. A category $\Sigma_{\mathbf{Alg}}$ of Scott topologies on algebraic lattices is readily obtained by employing continuous maps between topologies as morphisms.

**Theorem 6.13** The categories **Alg** and $\Sigma_{\mathbf{Alg}}$ are isomorphic, hence equivalent.

**Proof.** The required functors are defined on objects by taking the Scott topology and the specialization order of the arguments, respectively. By order consistency of the topologies, this yields a bijection between the classes of objects. Since the carrier sets of lattices and topologies remain unchanged, one can consider every function between algebraic lattices directly as a function between spaces and vice versa. To finish the proof, one needs to show that a function between algebraic lattices is Scott continuous iff it is continuous with respect to the Scott topologies. This standard result can for example be found in [AJ94]. □

In the next section, we see that the topological spaces of $\Sigma_{\mathbf{Alg}}$ are indeed very specific.

## 6.3 Stone duality

Since the very beginning of the theory, Stone duality has been recognized as a tool for relating proof theory, algebraic semantics, and model theory of logical systems (see [Sto37]). One direction of this investigation has already been mentioned in Section 6.1: Lindenbaum algebras can be represented by corresponding model theories, where models are characterized as subsets (filters) of formulae. Dually, one could also have presented every formula by the set of its models. The conceptual step from such systems of specific subsets to *topological spaces* was the key to the strength and utility of Stone's original representation theorems.

However, it still took decades to recognize that it would be even more advantageous to undo this step to the spacial side of Stone duality and to return to the more abstract world of partially ordered sets. It became apparent that topologies could not only serve as a representation for specific ordered structures, but that conversely orders could serve as a general substitute for topological spaces. Indeed, the leap to the spacial side is usually not an easy one — in many cases it cannot be made within classical Zermelo-Fraenkel set theory (ZF). The localic side on the other hand can mimic most of the features of the original topological setting, while being freed from the weight of points which often prevent purely constructive reasoning.

In what follows we embed our specific scenery into the setting of Stone duality. However, it turns out that the special case we consider does not justify to present the theory in its common generality. Hence we give explicit proofs for the object level relationships in our specialized setting and hint at the connections to more abstract versions of Stone duality where appropriate. Other than providing the merit of a more self-contained presentation, this also enables us to work exclusively in ZF, with no additional choice principles whatsoever. As a general reference on Stone duality, we recommend [Joh82].

The passage from spaces to orders is a particularly simple one: the open set lattice of a topology is already a poset. The class of posets arising in this way are the *spacial locales*.

**Definition 6.14** A complete lattice $L$ is a *locale* if the following infinite distributive law holds for all $S \subseteq L$ and $x \in L$:

$$x \wedge \bigvee S = \bigvee \{x \wedge s \mid s \in S\}.$$

A *point of a locale* is a principal prime ideal of $L$, i.e. a subset $p \subseteq L$ such that $p = \downarrow \bigwedge p$ and, for any $x \wedge y \in p$, $x \in p$ or $y \in p$. The set of all points of $L$ is denoted by $\mathsf{pt}(L)$.

A locale is *spacial* if, for any two elements $x, y \in L$ with $x \not\leq y$, there is a point $p \subseteq L$ such that $x \in p$ and $y \notin p$. $L$ is *spectral* if $L$ is algebraic, its greatest element is compact, and the meet of any two compact elements of $L$ is compact.

We remark that locales are also called *frames*, and that structures of this kind are equivalently characterized as *complete Heyting algebras*.[7]

It is now easy to see that any open set lattice yields a locale, where distributivity follows from the corresponding distributivity of finite intersections over infinite unions. Furthermore, Proposition 6.12 (ii), (iii), and (iv) show that, for an algebraic lattice $L$, $(\sigma(L), \subseteq)$ is even a spectral locale. We shall find that these locales are even more specific than this.

Our starting point for investigating topologies were algebraic lattices, which we have earlier recognized as the model theories of deductive systems of CCP logics. The abstraction to (certain) spectral locales brings us back to proof theory. We now characterize the above locales by relating them to Lindenbaum algebras of CCP logic, and reobtain topological spaces from this data.

We consider arbitrary meet-semilattices with greatest element, knowing that they are up to isomorphism just the Lindenbaum algebras of CCP (Theorem 6.5). Furthermore, we already mentioned that the collections of all filters (the order-dual concepts of the ideals) of such semilattices are just the algebraic lattices, which follows immediately from Theorem 3.6. We can now give a characterization for the locale of Scott open sets of algebraic lattices:

---

[7]The interested reader will find definitions and treatment in [Joh82, GHK$^+$03].

**Theorem 6.15** Consider a meet-semilattice $S$ with greatest element and the corresponding algebraic lattice $(\mathsf{Flt}(S), \subseteq)$ of filters of $S$. The collection of lower sets of $S$, ordered by subset inclusion, is isomorphic to $\sigma(\mathsf{Flt}(S))$. Every Scott open set lattice of an algebraic lattice is of this form.

**Proof.** Theorem 3.6 shows the bijective correspondence between the elements of $S$ and the compacts of $\mathsf{Flt}(S)$, since $S$ is dually order-isomorphic to $\mathsf{K}(\mathsf{Flt}(S))$. Proposition 6.12 demonstrated that every Scott open set is characterized by the compact elements it contains. Now it is obvious that such sets of compacts correspond to upper sets in the join-semilattice of compacts, and thus to lower sets in its dual meet-semilattice. The other direction is also immediate from the according part of Theorem 3.6. □

Hence the spectral locales of the form $\sigma(L)$ for some algebraic lattice $L$ are more precisely characterized as the lower set topologies of meet-semilattices with greatest element, i.e. as the *Alexandrov topologies* of join-semilattices with least element. Note also that meets and joins within these locales are really given by the corresponding set operations. By $\sigma_{\mathbf{Alg}}$ we denote the category of all locales isomorphic to the collection of lower sets on some meet-semilattice with greatest element together with functions that preserve finite meets and arbitrary joins.[8]

Next we want to connect up with the common constructions of Stone duality.

**Lemma 6.16** Consider a meet-semilattice with greatest element $S$ and its locale of lower sets $\sigma$. Then the meet-prime elements of $\sigma$ are exactly the complements of the filters of $S$.

**Proof.** Let $F \subseteq S$ be a filter and set $A = S \setminus F \in \sigma$. Now assume there are lower sets $B_1, B_2 \in \sigma$ such that $B_1 \cap B_2 = A$. For a contradiction, assume that there are elements $b_1 \in B_1 \cap F$ and $b_2 \in B_2 \cap F$. Then $b_1 \wedge b_2 \in F$ and $b_1 \wedge b_2 \in B_1 \cap B_2 = A$ — a contradiction. Hence, one of $B_1, B_2$ contains just the elements of $A$ as required.

Conversely, let $A \in \sigma$ be meet-prime and consider the upper set $F = S \setminus A$. For any two elements $a, b \in F$ it is easy to see that $\downarrow a \cap \downarrow b = \downarrow(a \wedge b)$. Hence, if $a \wedge b \in A$ then $\downarrow a \cup A$ and $\downarrow b \cup A$ are elements of $\sigma$ with intersection $A$, which cannot be. Hence $a \wedge b \in F$ as required. □

This gives us all necessary information about the points of these locales (see Definition 6.14), since these were defined to be just the principal ideals generated by meet-prime elements. We can thus identify the set of points $\mathsf{pt}(\sigma)$ with the set

---

[8]This is of course rather a category of frames and frame homomorphisms than a category of locales (which would be described by its dual). We have chosen to trade some terminological precision for conciseness of the presentation.

of all meet-prime elements of $\sigma$.[9] Our insights allow us to give a direct description of the topological spaces associated with semilattices:

**Corollary 6.17** Let $S$ be a meet-semilattice with greatest element, let $L$ be an algebraic lattice, and let $\sigma$ be a spectral locale, such that

- $S^{op}$ is isomorphic to $\mathsf{K}(L)$ and
- $\sigma$ is isomorphic to $\sigma(L)$.

Then the following are homeomorphic:

(i) $(L, \sigma(L))$, the Scott topology on $L$;

(ii) the topology on $\mathsf{Flt}(S)$ generated from the basic open sets

$$O_a = \{F \in \mathsf{Flt}(S) \mid a \in F\} \qquad \text{for all } a \in S;$$

(iii) the topology on $\mathsf{pt}(\sigma)$ given by the open sets

$$P_A = \{p \in \mathsf{pt}(\sigma) \mid A \notin p\} \qquad \text{for all } A \in \sigma.$$

**Proof.** Most of the above should be obvious at this stage, so we spare out some details. Suitable bijections between $L$, $\mathsf{Flt}(S)$, and $\mathsf{pt}(\sigma)$ have been obtained in 3.6 and 6.16. First we show the homeomorphism between (i) and (ii) (which induces also that $(O_a)$ is indeed a base). For this we only have to note that $O_a = \{F \in \mathsf{Flt}(S) \mid \uparrow a \subseteq F\}$. Using the bijection between (principal) filters and (compact) elements from Theorem 3.6, one sees that $O_a$ corresponds to an open set $\uparrow c$, $c \in \mathsf{K}(L)$, of (i). The fact that these subsets are open and form the basis for the Scott topology has been shown in Proposition 6.12.

For the homeomorphism between (ii) and (iii), we consider the locale of lower sets of $S$, which is isomorphic to $\sigma$ by Theorem 6.15. Clearly this affects the topology of (iii) only up to homeomorphism. Now in the locale of lower sets, a point (principal prime ideal) $p = \downarrow B$ is in $P_A$ iff the corresponding meet-prime $B$ does not contain $A$. But this is the case iff the complement of $B$ intersects $A$. Hence, by Lemma 6.16, $P_A$ corresponds exactly to the collection of those filters of $S$ that contain some element of $A$, i.e. to the set $\bigcup\{O_a \mid a \in A\}$. But these are precisely the open sets of the topology of (ii). □

---

[9]Furthermore, we remark that this guarantees a sufficient supply of prime elements without invoking any additional choice principles, i.e. we are dealing with a class of locales that is spacial in Zermelo-Fraenkel set theory. This contrasts with the class of all spectral locales, which is only spacial when the existence of prime ideals is explicitly postulated, i.e. when the *Boolean prime ideal theorem* [DP02] is assumed to hold.

With respect to the given preconditions on the relationship between $S$, $L$, and $\sigma$, note that the various transformations between semilattices, algebraic lattices, and locales established earlier yield a variety of equivalent ways to state that the three given objects describe "the same thing".

To complete the targeted categorical equivalence between the dual category of $\sigma_{\mathbf{Alg}}$ and $\Sigma_{\mathbf{Alg}}$ (**Alg**, **Sem**$_\vee$, ...), one still needs to prove a suitable bijection of hom-sets. This correspondence between inverse frame homomorphisms and continuous functions is a basic result of Stone duality which we will not repeat here. See [Joh82] for details.

# 7 Summary and further results

We provided characterizations of the category of algebraic lattices by means of structures from logic, topology, domain theory, and formal concept analysis. More precisely, we characterized algebraic lattices by certain semilattices, formal contexts, and deductive systems of the conjunctive fragment of propositional logic. The novel category **Cxt** of formal contexts and approximable mappings was used to establish the cartesian closure of these categories, and the categorical constructions needed for this were explicitly given. Other representations referred to special classes of closure systems, Scott topologies, locales, and definite logic programs. An overview of the major equivalences given herein is displayed in Figure 2.

Although this treatment is quite comprehensive, one could still add some more equivalent formalisms. Especially, we left out the *coverage technique* of [Joh82] (see also [Sim04]), which represents locales in a syntactical way that relates closely to Scott information systems. Furthermore, we deliberately ignored Scott's earlier approach to presenting domains via *neighborhood systems* [Sco82b], since these structures are not much more than a mixture of the later (token-set based) information systems and continuous closure operators. Finally, one could also identify the classes of distributive lattices that arise as the compact elements of the spectral locales we considered as the free distributive lattices over the underlying semilattice.

In this article we have also presented a unified treatment of the basic techniques and mechanisms that are applied to join domain theory, algebra, logic, and topology. Algebraic lattices turn out to be the simplest case where such a discussion is feasible. Part of the given results have been generalized in various ways, some of which are subject to current research. A common way to generalize the above results is to extend the logic under consideration. A technique for including "negation-like" constraints without need for an internal negation operation has been employed by Scott in his original formulation of information systems [Sco82a]. There he introduced a collection of finite subsets of propositions that are consistent, assuming that no inconsistent sets can be mapped to true by any model. This procedure can be viewed as an extension of the deductive system that allows statements of the form "$X \Vdash$", interpreted as $\bigwedge X \Vdash \bot$, where $\bot$ is the constant *false* — a construction well-known under the notion of *integrity constraint* in database theory. Clearly, $\bot$ will then represent the least element in the resulting Lindenbaum algebras. However, as important as introducing the constant $\bot$ into the logical language is a change of the model theory: models now have to be *proper* filters, i.e. the case that all (including $\bot$) formulae evaluate to true is excluded. The posets of models for such logics turn out to be exactly the Scott domains (the bounded complete algebraic cpos).

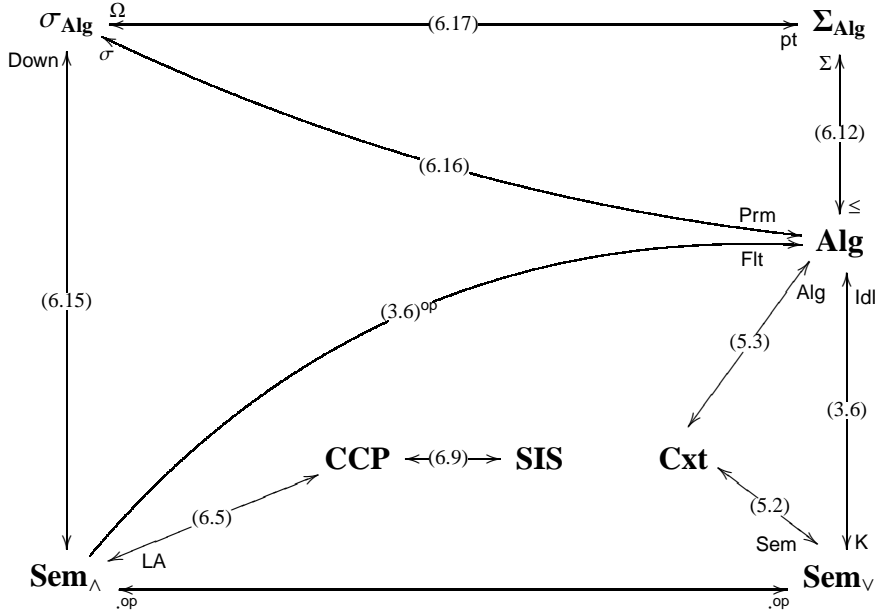

Figure 2: Summary of all established equivalences with reference to the corresponding (object-level) statements. Labels at the arrow tips specify the name of the functor that was used in a construction, where Down denotes the construction of the lower set topology from a meet-semilattice, Prm yields the set of principal prime ideals of a locale, ordered by subset inclusion, and $\leq$ denotes the construction of the specialization order from a topological space.

As another step, one can include disjunction into the formalism. This already leads to a substantial complication of the theory: choice principles are now needed to obtain models. Since logical conjunction and disjunction are classically assumed to distribute over each other, one obtains all (bounded) distributive lattices as Lindenbaum algebras. In place of algebraic lattices one finds a curious class of dcpos that have been termed *information domains* in [DG90]. Later the direct construction of distributive lattices and locales from such deductive systems was studied in [CC00] and [CZ00], and in [RZ01] Smyth powerdomains were characterized by similar means using a clausal logic which was also extended to non-monotonic reasoning paradigms on hierarchical knowledge [RZ01, Hit04]. Other than this, one can apply all the representation machinery that has been set up for distributive lattices, including both Stone's and Priestley's representation theorems for these structures ([Joh82]).

Further strengthening of the logic is possible by including some internal negation operation. Intuitionistic negation yields Heyting algebras as Lindenbaum algebras. The resulting topologies are already studied in [Sto37], though the significance of specialization orders and domain theoretic concepts were not yet recognized at this time. If classical negation is preferred instead, thus yielding classical propositional logic, the class of Boolean algebras provides the well-known algebraic semantics. While topological representation via Stone's theorem is rather pleasant in this case, the domain theoretic aspects are quite disappointing: the specialization order of models is discrete. Related approaches nevertheless have been taken for the context of negation in logic programming [Sed95, Hit04], but the domain-theoretic content of these investigations remains to be determined.

For reasons as those just described, internal negation is usually not considered in domain-theoretical studies. However both inconsistency of finite subsets and finite disjunctions can be employed with various restrictions to obtain classes of domains that are more general than the Scott domains. A slight constraint on either the logical ([DG90]) or the localic level ([Abr91]) restricts the obtained class of dcpos (of models) to the coherent algebraic dcpos. However, while this is a well-known concept in domain theory, it results in rather unusual restrictions on the logics (Lindenbaum algebras, locales). Further conditions will lead to SFP-domains [Abr91, Zha91]. On the other hand, conditions that characterize a class of deductive systems that produces exactly the L-domains have been studied in [Zha92].